\documentclass[aps,pra,twocolumn ,groupedaddress]{revtex4-1}
\usepackage{graphicx}
\usepackage[final]{changes}
\usepackage{amsmath}
\usepackage{float}
\usepackage{hyperref}
\usepackage{color}

\begin{document}


\title{Bath-induced decay of Stark many-body localization}


\author{Ling-Na Wu}
\email[]{lnwu@pks.mpg.de}
\affiliation{Max Planck Institute for the Physics of Complex Systems,
D-01187, Dresden }

\author{Andr{\'e} Eckardt}
\email[]{eckardt@pks.mpg.de}
\affiliation{Max Planck Institute for the Physics of Complex Systems,
D-01187, Dresden }


\date{\today}

\begin{abstract}
We investigate the relaxation dynamics of an interacting Stark-localized system coupled to a dephasing bath, and compare its behavior to the conventional disorder-induced many body localized system. Specifically, we study the dynamics of population imbalance between even and odd sites, and the growth of the von Neumann entropy.
For a large potential gradient, the  imbalance is found to decay  on a time scale $\tau$ that grows quadratically with the Wannier-Stark tilt.
\added{For the non-interacting system, it shows an exponential decay, which becomes a stretched exponential decay in the presence of finite interactions.}
This is different from a system with disorder-induced localization, where the imbalance exhibits a stretched exponential decay \added{also for vanishing interactions}.
As another clear qualitative difference, we do not find a logarithmically slow growth of the von-Neumann entropy as it is found for the disordered system.
Our findings can immediately be tested experimentally with ultracold atoms in optical lattices.
\end{abstract}


\maketitle

Many body localization (MBL)~\cite{Altman2015MBLReview, Nandkishore2015MBLReview, ALET2018MBLReview,abanin2018ergodicity}, which describes the failing of an interacting system with quenched disorder to thermalize, has attracted widespread attentions in recent years, \added{both theoretically~\cite{Peter2008PhysRevB.77.064426,Bardarson2012PhysRevLett.109.017202,Serbyn2013PhysRevLett.111.127201,Huse2014PhysRevB.90.174202,Chandran2015PhysRevB.91.085425,Luitz2015PhysRevB.91.081103,Serbyn2015PhysRevX.5.041047}  and experimentally~\cite{Schreiber2015Science,smith2016NP,Choi2016Science,Roushan2017Science,2018arXiv181206959R}}. Over the
past decade, studies have uncovered a rich variety of unique and interesting properties of MBL phases, such as logarithmic growth of entanglement~\cite{Peter2008PhysRevB.77.064426,Bardarson2012PhysRevLett.109.017202}, the emergence
of an extensive set of quasi-local integrals of motion~\cite{Serbyn2013PhysRevLett.111.127201,Huse2014PhysRevB.90.174202,Chandran2015PhysRevB.91.085425}, the existence of many-body mobility edges~\cite{Luitz2015PhysRevB.91.081103,Serbyn2015PhysRevX.5.041047}, and so on.

So far, most of the studies on MBL are based on disordered system. However, it is a very intriguing question whether MBL can be achieved also without disorder.
The idea of disorder-free localization
can be traced back to the early work on interaction-induced localization~\cite{kagan1984localization}. A lot of efforts have been devoted to the possibility of MBL in translation-invariant systems~\cite{Grover2014JSM,PhysRevB.91.184202,schiulaz2014ideal,PhysRevLett.117.240601,PAPIC2015714,PhysRevLett.118.266601,PhysRevLett.119.176601,DeRoeck2014,Hickey2016JSM,PhysRevB.92.100305,carleo2012localization}. Most of them are based on the mixture of two species of particles~\cite{Grover2014JSM,PhysRevB.91.184202,schiulaz2014ideal,PAPIC2015714,PhysRevLett.117.240601}, where
one species effectively acts as a disorder potential. \added{However, a recent study~\cite{PAPIC2015714} concludes that these models show only transient localized behavior, which ultimately becomes
delocalized at long times.}
Recently, two papers~\cite{van2018bloch,schulz2018stark} explored another direction by looking for MBL-like behavior in interacting Wannier-Stark localized systems.
\added{These models are shown to exhibit nonergodic behavior as indicated by their spectral and dynamical properties.}

In search for evidence of MBL in the absence of disorder, all the previous studies focus on closed (isolated) systems and show several hallmarks of MBL in their models.
On the other hand, in recent years, the imperfect experimental environment has excited an
intense interest in the effect of dissipation on MBL~\cite{Levi2016PhysRevLett.116.237203,Fischer2016PhysRevLett.116.160401,Medvedyeva2016PhysRevB.93.094205,Everest2017PhysRevB.95.024310,PhysRevB.76.052203,Bloch2017PhysRevX.7.011034,Nandkishore2014PhysRevB.90.064203,Nandkishore2016AP,Rubio2018arXiv,2018arXiv180604772L,Johri2015PhysRevLett.114.117401,Nandkishore2015PhysRevB.92.245141,Luitz2017PhysRevLett.119.150602,Hyatt2017PhysRevB.95.035132,Wu2018arXiv181106000}. When the system is coupled to environments with broad spectrum, the MBL phase will eventually be destroyed. However, the relaxation can be extremely slow in the open disorder-induced MBL systems~\cite{Levi2016PhysRevLett.116.237203,Fischer2016PhysRevLett.116.160401}.

Here, we explore the fate of disorder-free localization  in the presence of dissipation. Specifically, we study the Wannier-Stark localized system~\cite{van2018bloch,schulz2018stark} coupled to a dephasing bath.
This type of dissipation, which has been studied in a number of recent papers~\cite{Levi2016PhysRevLett.116.237203,Fischer2016PhysRevLett.116.160401,Medvedyeva2016PhysRevB.93.094205,Everest2017PhysRevB.95.024310,Marko2016AP},
is particularly relevant for experiments with ultracold atoms in optical lattices, where it is induced by the off-resonant scattering of lattice photons via spontaneous emission~\cite{Zoller2010PhysRevA.82.063605,Bloch2017PhysRevX.7.011034}.

Starting from a density-wave state with one fermion on every other site of a one-dimensional lattice, we investigate
the dynamics of
population imbalance between even and odd sites, and the growth of von Neumann entropy.
In the limit of strong localization, the relaxation dynamics is found to become very slow and dependent on the field gradient.
However, the way both entropy and imbalance relax is found to be qualitatively different from the case of disorder-induced MBL systems~\cite{Levi2016PhysRevLett.116.237203,Fischer2016PhysRevLett.116.160401}.


\begin{figure*}[!htp]
    \centering
      {\includegraphics[width=0.67\columnwidth]{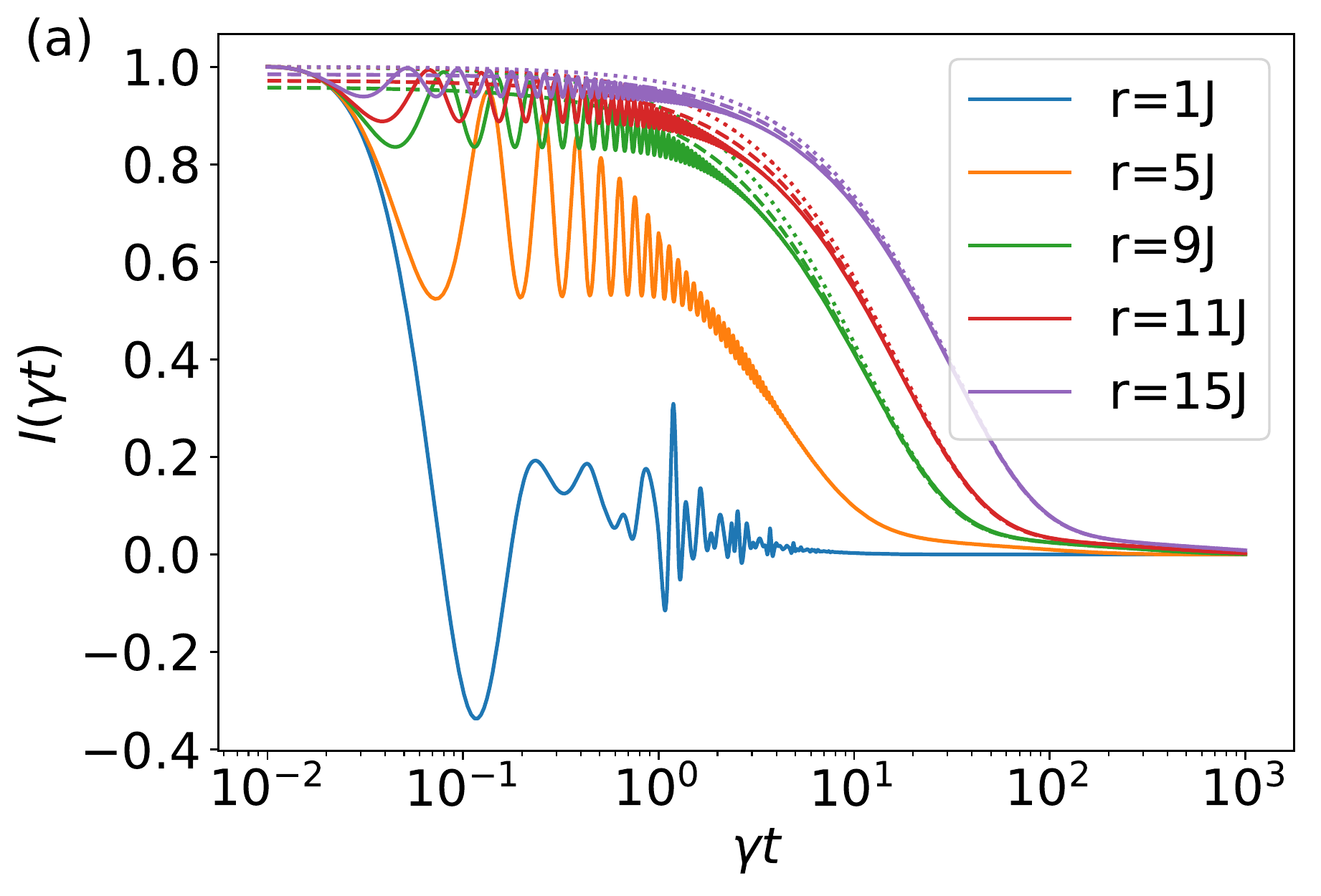}}
     {\includegraphics[width=0.67\columnwidth]{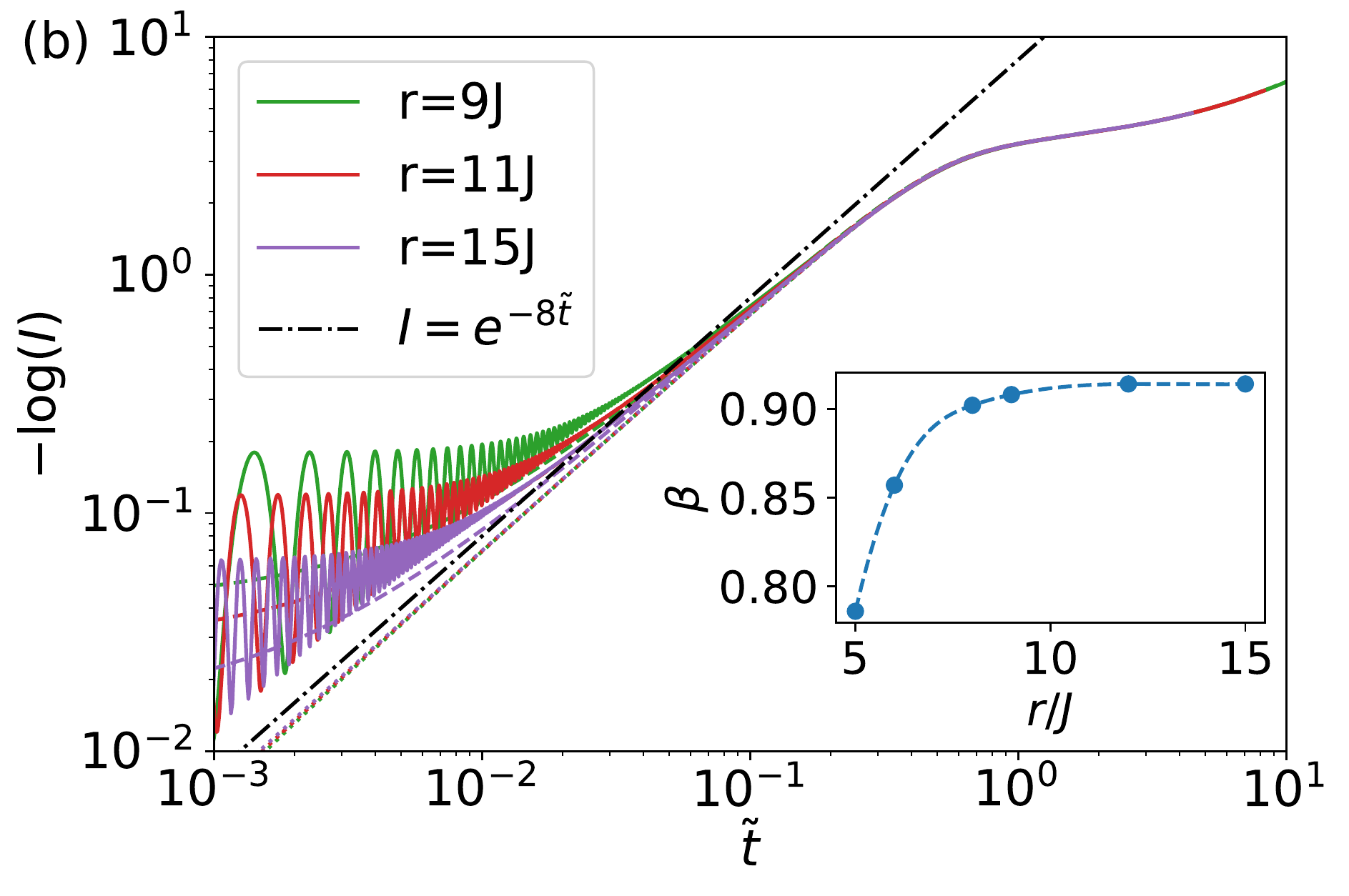}}
      {\includegraphics[width=0.67\columnwidth]{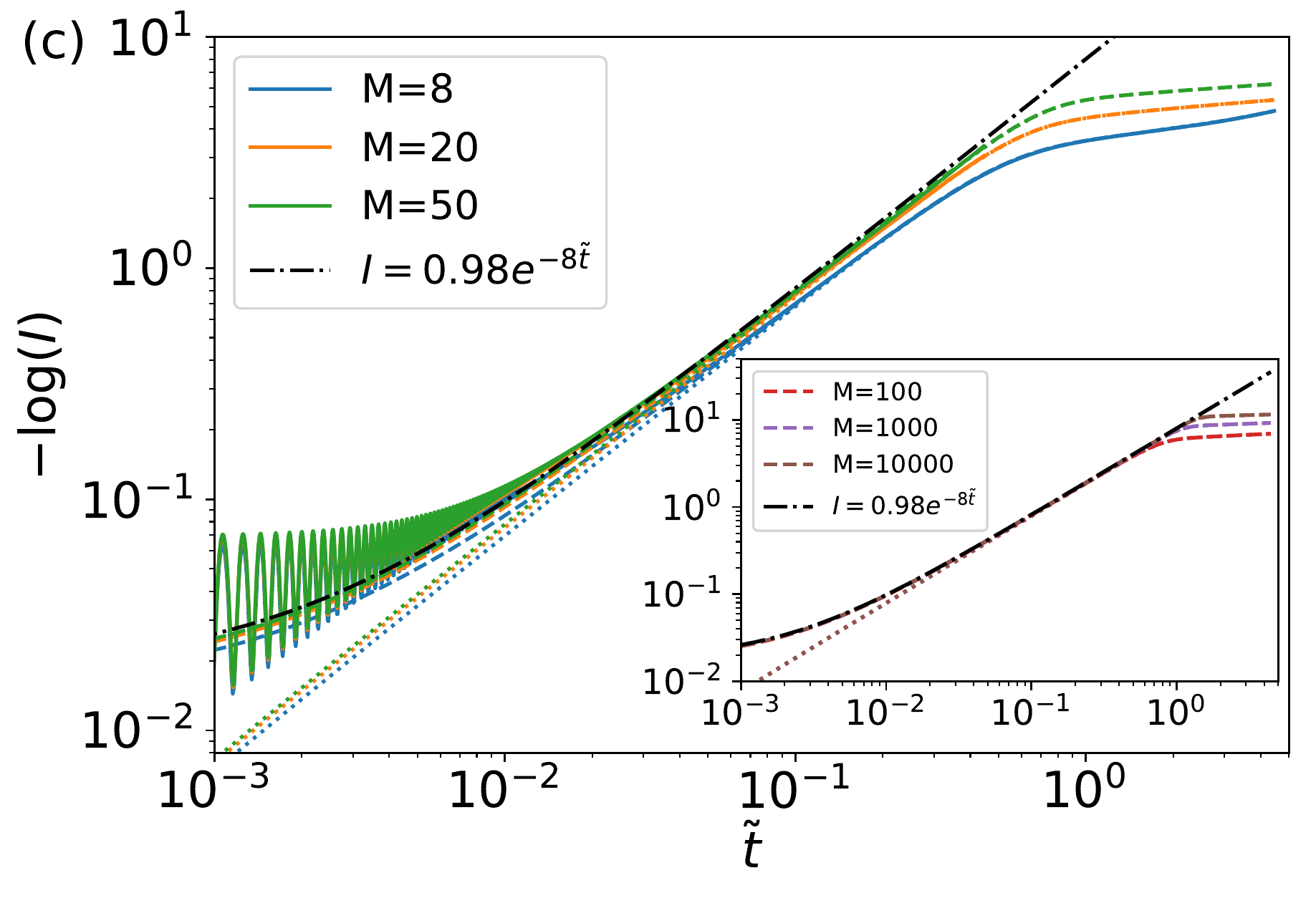}}
    \caption{Dynamics of the population imbalance between even and odd sites, $I$, for non-interacting system from the initial charge-density-wave state. (a) $I$ as a function of $\gamma t$ for $M=8$. (b) $-\log(I)$ as a function of the scaled time $\tilde{t} = \gamma t (J/r)^2$ for $M=8$.  The black dot-dashed line denotes the simple expression given in Eq.~\eqref{Iapp}. \added{The inset shows the stretching exponent $\beta$ of the fitting curve $I=I_0 e^{-(t/t_0)^\beta}$ as a function of field gradient $r$. The dashed line is a guide to the eye.} (c) $-\log(I)$ at $r=15J$ for different system sizes $M$. The black dot-dashed lines are fitting curves based on exponential decay function. {For all the plots, the solid lines denote the results obtained by numerical integration of the master equation~\eqref{L} ($M=8$) or by time evolving a density matrix in tensor-product form with Trotter Gates using the ITensor library~\cite{ITensor} ($M \ge 20$). The dashed (dotted) lines depict the approximated imbalance $I$ ($\tilde I$) from the classical rate equations~\eqref{nk1}, which overlap with ($M=8$) or smoothly connect to ($M\ge20$) the solid lines at late times.}
    The dissipation rate is $\gamma=0.1J$.
    }\label{It}
\end{figure*}

The model under consideration is a chain of interacting spinless
fermions with open boundary conditions, subject to a strong electric field, with Hamiltonian
\begin{eqnarray}\label{H}
H &=& J\sum\limits_{i=1}^{M-1}{\left(c_i^\dag c_{i+1} + c_{i+1}^\dag c_i \right)} + \sum\limits_{i=1}^{M}{W_i n_i}+V\sum\limits_{i=1}^{M-1}{n_i n_{i+1}}.\notag\\
\end{eqnarray}
Here the operator $c_i^\dag$ creates a fermion on lattice site $i$, and $n_i = c_i^\dag c_i$ is the associated number operator.
The first term in (\ref{H}) denotes tunneling between nearest neighbor sites with rate $J$. The second term is the on-site potential describing the applied static gradient, $W_i= -r i$ .
The last term describes the nearest-neighbor interactions with strength $V$.

The  non-interacting system (in the thermodynamic limit) exhibits the well-known Wannier-Stark effect~\cite{PhysRev.117.432}, where the particles are localized due to the
linear potential. In the Wannier representation, the single-particle eigenstates~\cite{Fukuyama1973PhysRevB.8.5579} take the form $|k\rangle_\infty = \sum_i J_{i-k}(\lambda)|i\rangle$, where  $J_{n}(\lambda)$ is  the Bessel function of the first kind with argument $\lambda=-2J/r$. The associated eigenenergies $E_k = -rk$ form the Wannier-Stark ladder with equal level splittings determined by the electric field.
When interactions are turned on, the system is shown to remain localized above a critical potential gradient $r$ and to
exhibit non-ergodic behavior analogous to conventional MBL~\cite{van2018bloch,schulz2018stark}, such as logarithmic growth of entanglement entropy,  Poissonian level statistics, etc.~\footnote{Strictly speaking, the system with purely linear potential shows singular behavior~\cite{schulz2018stark} and a small perturbation, e.g.\ in the form of a quadratic potential, has to be added to recover generic MBL behavior. However, below we show that adding such a perturbation does not lead to a qualitative change of the dynamics of the open system.}.

We couple the system to a dephasing bath that couples to the on-site occupations and
which can be interpreted as a structureless environment
allowing for energy exchange at all scales. \added{Such a dissipation can be engineered in experiments with ultracold atoms via the off-resonant scattering of lattice photons~\cite{Zoller2010PhysRevA.82.063605,Bloch2017PhysRevX.7.011034}.} It drives the system towards infinite-temperature state in the long-time limit.
The full dynamics of the system
can be described by a master equation of
Lindblad form~\cite{BreuerPetruccione},
\begin{equation}\label{L}
\frac{d \rho}{dt} = -i\left[H,\rho\right] + \gamma \sum\limits_{i=1}^{M}{\left(n_i \rho n_i - \frac{1}{2}n_i^2\rho -  \frac{1}{2}\rho n_i^2\right)},
\end{equation}
where $\rho$ is the system's density matrix and $\gamma>0$ sets the
coupling to the bath.

In order to study ergodicity breaking in the open system, let us first investigate the dynamics of population imbalance between even and odd sites,
\begin{equation}
I(t) = \frac{N_{\rm even}(t) - N_{\rm odd}(t)}{N}.
\end{equation}
Being easily accessible, this quantity is widely used in experiments~\cite{Schreiber2015Science,Bordia2016PhysRevLett.116.140401,Choi2016Science,Bordia2017PhysRevX.7.041047,Bloch2017PhysRevX.7.011034} to quantify the memory of the initial conditions.
We choose a charge-density-wave state with every second lattice site occupied as an initial state, which is also the usual choice in experiments.
For the isolated system in the MBL phase, the imbalance approaches a finite value in the steady state~\cite{van2018bloch,schulz2018stark}. Such a memory of the initial condition can no longer be maintained in the presence of
dissipation.


Figure~\ref{It}(a) shows the dynamics of the imbalance $I$ for non-interacting systems ($M=8$) with different field gradients $r$. The imbalances (solid lines) oscillate at short times $ t <1/\gamma$ and then decay to zero at a rate that is found to depend on the field gradient $r$.
We show $-\log(I)$ for large $r$ in a logarithmic plot as a function of the scaled time $\tilde{t}= t/\tau$ with
\begin{equation}\label{tau}
\tau = \gamma^{-1} (r/J)^2
\end{equation}
in Fig.~\ref{It}(b). The results (solid lines) collapse onto each other from the time where decay sets in.
\added{The decay is found to be approximately exponential.
By fitting it to a stretched exponential function $I=I_0 e^{-(t/t_0)^\beta}$, we get a stretching exponent $\beta \simeq 0.9$ close to $1$ at a large field gradient $r$, as shown in the inset of Fig.~\ref{It}(b) [see Fig. S2 in Supplementary Material for more details of the curve fitting]. The slightly stretched exponential behavior is a finite-size effect: it approaches an exponential decay as system size increases, as shown in Fig.~\ref{It}(c).} This decay behavior is different from that for disorder-induced localization~\cite{Fischer2016PhysRevLett.116.160401}. There the population imbalance exhibits a stretched exponential decay \added{with stretching exponent $\beta \simeq 0.38$} for the non-interacting system under dephasing noise. This behavior is attributed to the different decay rates at distinct parts of the system
due to fluctuations in the disorder strength, which are absent in our model.

To explain the observed behavior in our model, let us study the dynamics of the mean occupations in the eigenbasis.
The dephasing noise leads to the decay of the off-diagonal elements of the density matrix in the eigenstate basis due to rapid oscillations,
resulting in a diagonal density matrix for long-time evolution~\cite{Fischer2016PhysRevLett.116.160401,Wu2018arXiv181106000,DanielPRE}. Hence, the equation of motion for the mean occupation of the eigenstate $\langle {\tilde n}_k \rangle$ can be well described by the following classical rate equation
\begin{equation}\label{nk}
\langle \dot{{\tilde n}}_k \rangle = \gamma \sum\limits_{q}{\left(R_{kq} \langle {{\tilde n}}_q \rangle-R_{qk} \langle {{\tilde n}}_k \rangle \right)},
\end{equation}
where the jump rate
$R_{kq} = \sum_i{|\psi_{ik}^*\psi_{iq}|^2} = R_{qk}$ depends on the overlap of the two involved single-particle wavefunctions, $|k\rangle = \sum_i\psi_{ik} |i\rangle$, $|q\rangle = \sum_i \psi_{iq}|i\rangle$.

For a strong field, $r \gg J$, we take the Wannier-Stark states for the infinite system as the eigenstates for the finite-size system considered here, i.e., $\psi_{ik} = J_{i-k}(\lambda)$, which turns out to be a good approximation as shown later.
Due to the strong localization of the eigenstates, the rate in Eq.~\eqref{nk} is dominated by
$R_1 \equiv R_{k,k\pm1} \simeq 2 J_0(\lambda)^2 J_1(\lambda)^2 \simeq \lambda^2/2$, which connects nearest neighbors.
Hence,  Eq.~(\ref{nk}) can be reduced to
\begin{eqnarray}\label{nk1}
\langle \dot{{\tilde n}}_1 \rangle &=& \gamma R_1 \left(\langle \tilde{n}_{2} \rangle - \langle \tilde{n}_{1} \rangle\right), \notag\\
\langle \dot{{\tilde n}}_k \rangle &=& \gamma R_1 \left(\langle \tilde{n}_{k+1} \rangle + \langle \tilde{n}_{k-1} \rangle -2 \langle \tilde{n}_{k} \rangle\right), {\text{ for}} \, 1<k<M \notag\\
\langle \dot{{\tilde n}}_M \rangle &=& \gamma R_1 \left( \langle \tilde{n}_{M-1} \rangle - \langle \tilde{n}_{M} \rangle\right),
\end{eqnarray}
whose explicit solution is given in the Supplemental Material.
Note that to obtain the population imbalance, we need the mean occupation in real space $\langle n_i \rangle$. While for a large $r$, we have
$\langle n_i \rangle = \sum_k{J_{i-k}(\lambda)^2 \langle \tilde{n}_k \rangle} \simeq \langle \tilde{n}_i \rangle$.
{The resulting population imbalance from $\langle n_i\rangle $ ($\langle \tilde{n}_i \rangle$) is shown as dashed (dotted) lines in Fig.~\ref{It}.}
Of course this approximation is not able to capture
the short-time oscillations due to the neglect of off-diagonal terms in the density matrix. Nevertheless,
for the long-time evolution, it agrees well with the  exact solution (solid lines)
obtained by numerical
integration of the master equation (\ref{L}) (for system with $M=8$ sites).

In order to get a simple expression for the imbalance, we make a further approximation.
From~Eq.~(\ref{nk1}), we can obtain the time evolution of the population in even and odd sites,
$\dot {{\tilde N}}_{\rm even} = -\dot {{\tilde N}}_{\rm odd} = 2\gamma R_1( {{\tilde N}}_{\rm odd} - {{\tilde N}}_{\rm even} ) -\gamma R_1 \left(\langle \tilde{n}_{1} \rangle - \langle \tilde{n}_{M} \rangle\right)$.
Thus, the dynamics of the population imbalance is governed by
\begin{eqnarray}\label{Ia1}
\dot{\tilde{I}} \equiv \frac{\dot {{\tilde N}}_{\rm even}  -  \dot {{\tilde N}}_{\rm odd}}{N} &=& -4\gamma R_1 \tilde{I} -2\gamma R_1 \frac{\langle \tilde{n}_{1} \rangle - \langle \tilde{n}_{M} \rangle}{N}.
\end{eqnarray}
By neglecting the edge term $\propto \left(\langle \tilde{n}_{1} \rangle - \langle \tilde{n}_{M} \rangle\right)/N$, we arrive at
\begin{equation}\label{Iapp}
\tilde{I} \simeq \tilde{I}(0) e^{-4 \gamma R_1 t} = \tilde{I}(0) e^{-8 \tilde{t}} = \tilde{I}(0) e^{-8 t/\tau}.
\end{equation}
This simple expression is shown by the black dot-dashed line in Fig.~\ref{It}(b).
It explains the observed approximately exponential decay of the imbalance on the time scale $\tau$ [Eq.~\eqref{tau}].
\added{By comparing it with the results from Eq.~\eqref{Ia1} (dotted lines, which overlap with the solid lines at long times),
we can see that the edge term leads to the deviation from the exponential decay for $\tilde t > 1$.
This is further confirmed in Fig.~\ref{It}(c), where the edge effect becomes weaker for a larger system.}

\begin{figure}[!htbp]
    \centering
{\includegraphics[width=0.47\columnwidth]{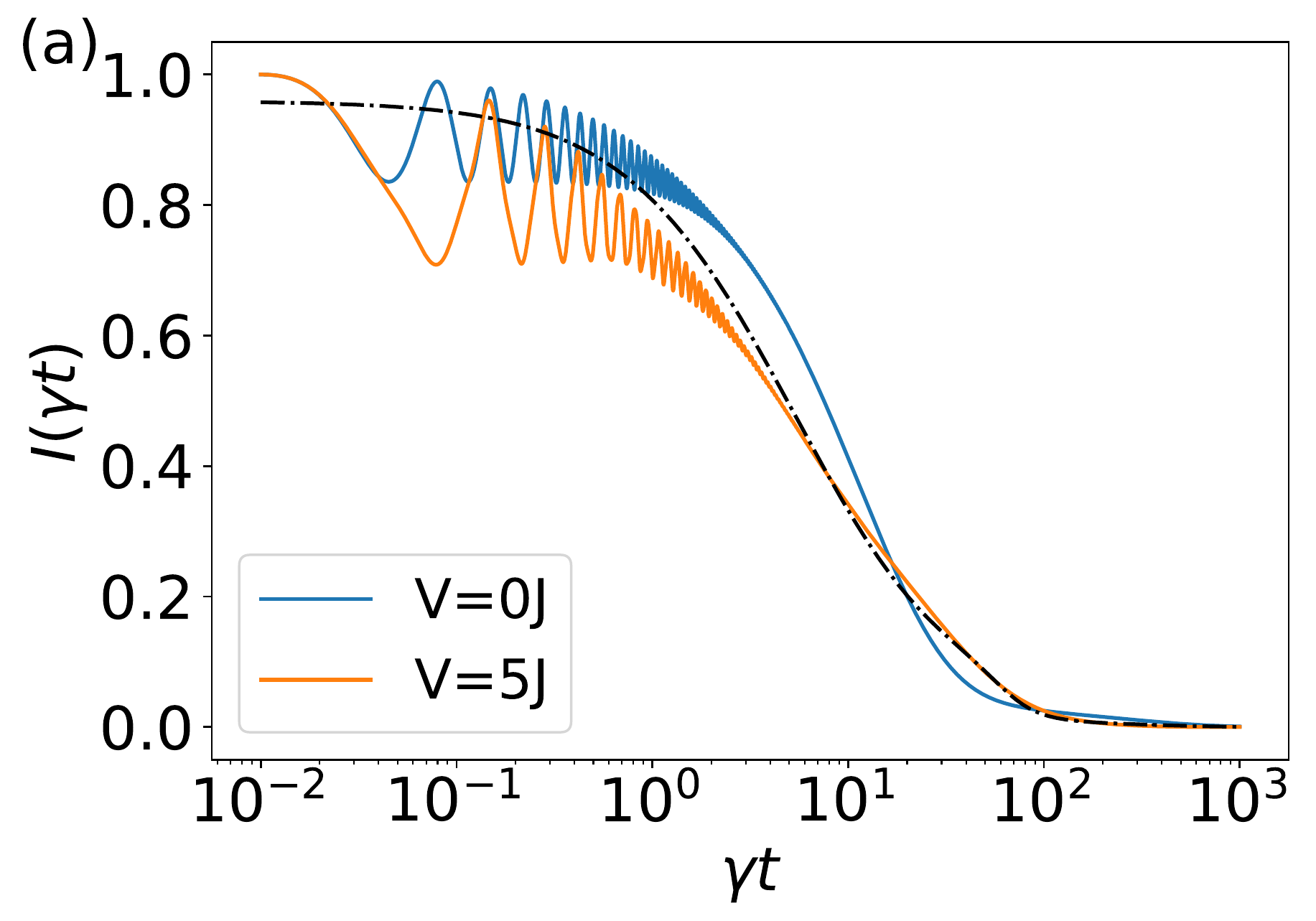}}
{\includegraphics[width=0.5\columnwidth]{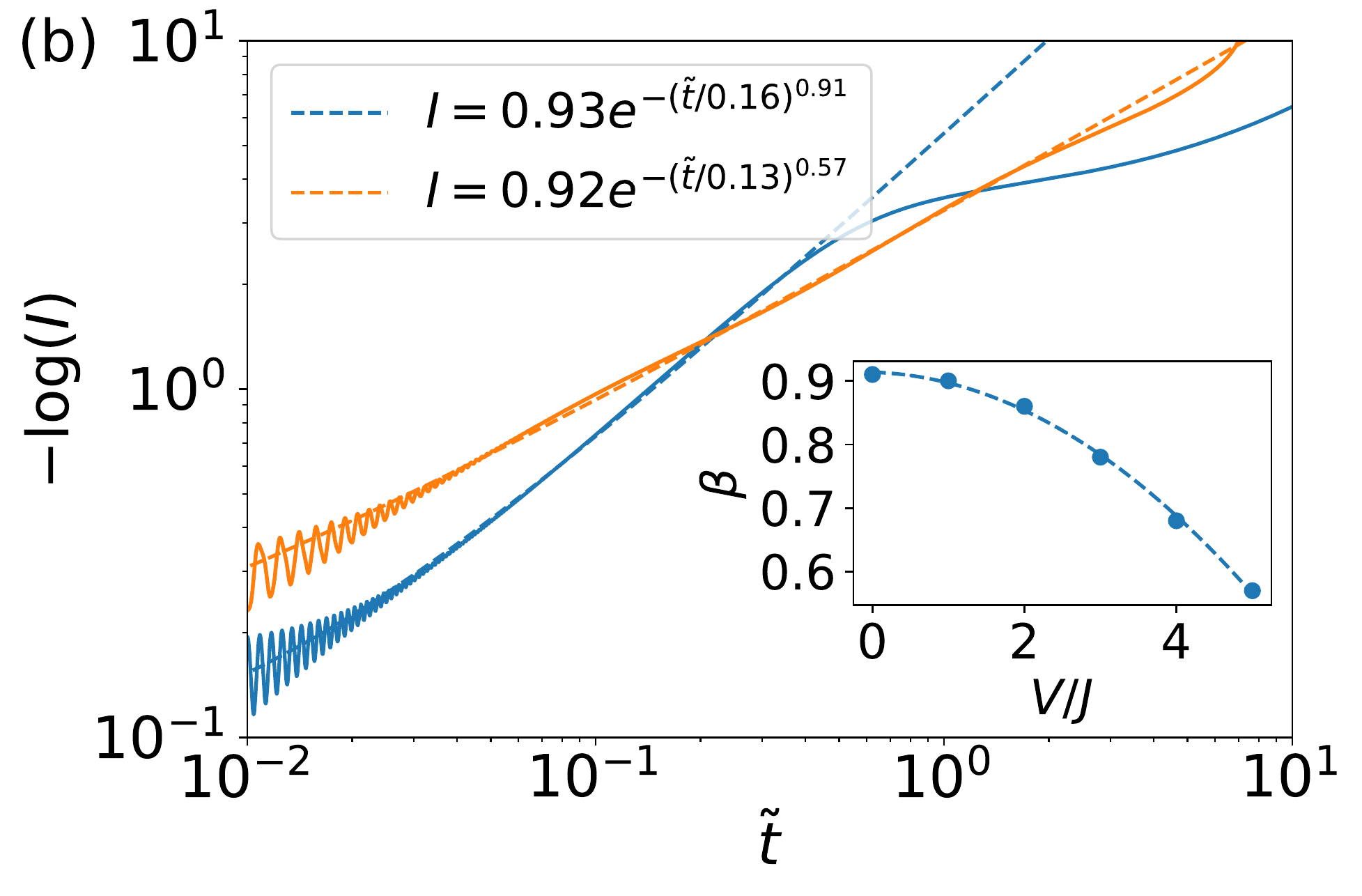}}
    \caption{Dynamics of the imbalance for interacting systems (orange solid line) from the initial charge-density-wave state.
    As a comparison, the result for non-interacting case is shown as blue line. The black dot-dashed line in (a) is the approximated result obtained from
the simple equations in \eqref{nkV} by using $\langle n_i\rangle = \sum_k |\psi_{ik}|^2 \langle \tilde{n}_k \rangle$. \added{The dashed lines in (b) are fitting curves based on a stretched exponential function $I_0 e^{-(t/t_0)^\beta}$. The inset shows the stretching exponent $\beta$ as a function of interaction strength $V$.} 
    The parameters are  $M=8$, $\gamma=0.1J$, $r=9J$, $\epsilon = 2.2$.
    }\label{IVt}
\end{figure}

 \begin{figure*}[!htbp]
    \centering
{\includegraphics[width=0.67\columnwidth]{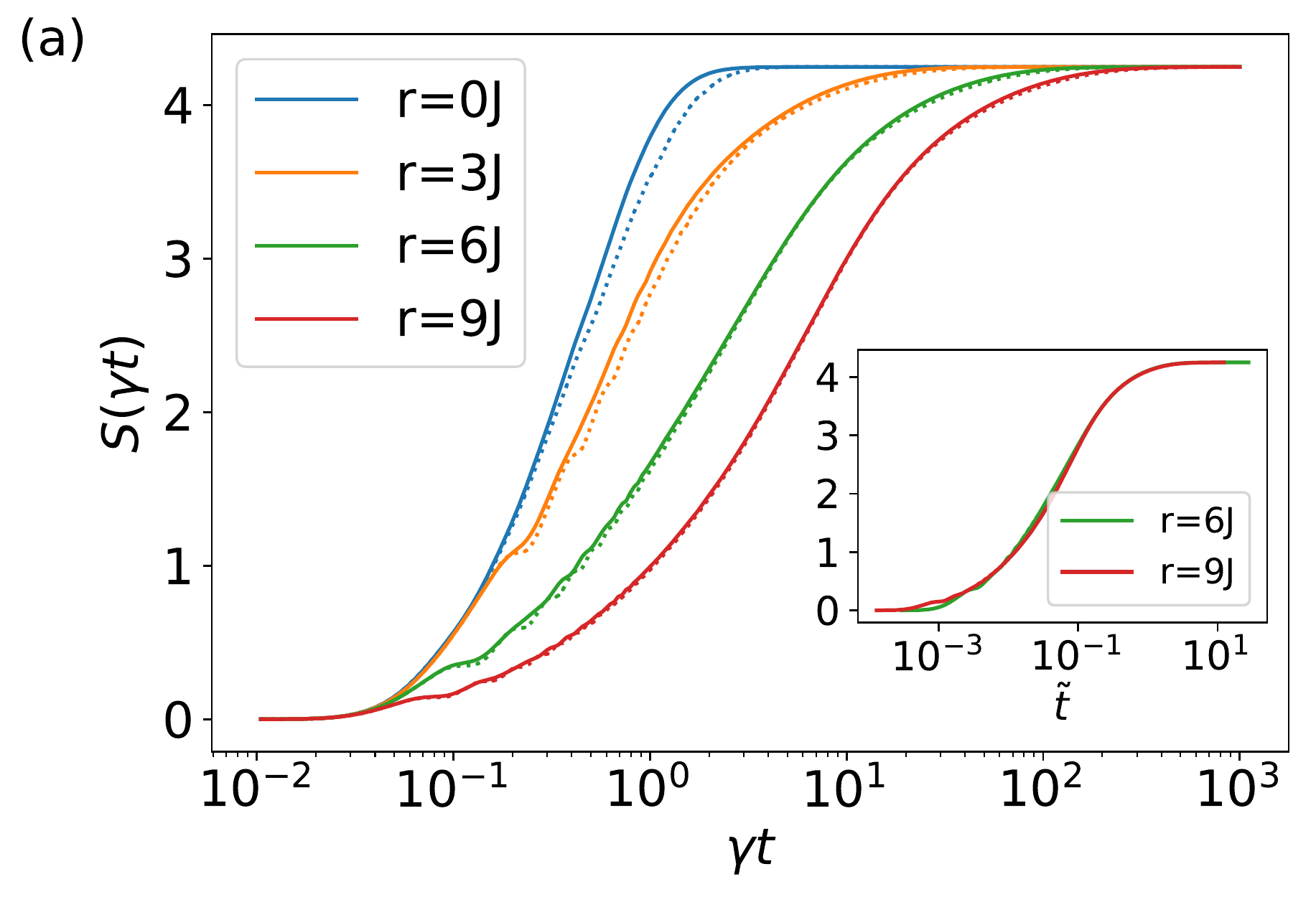}}
          {\includegraphics[width=0.67\columnwidth]{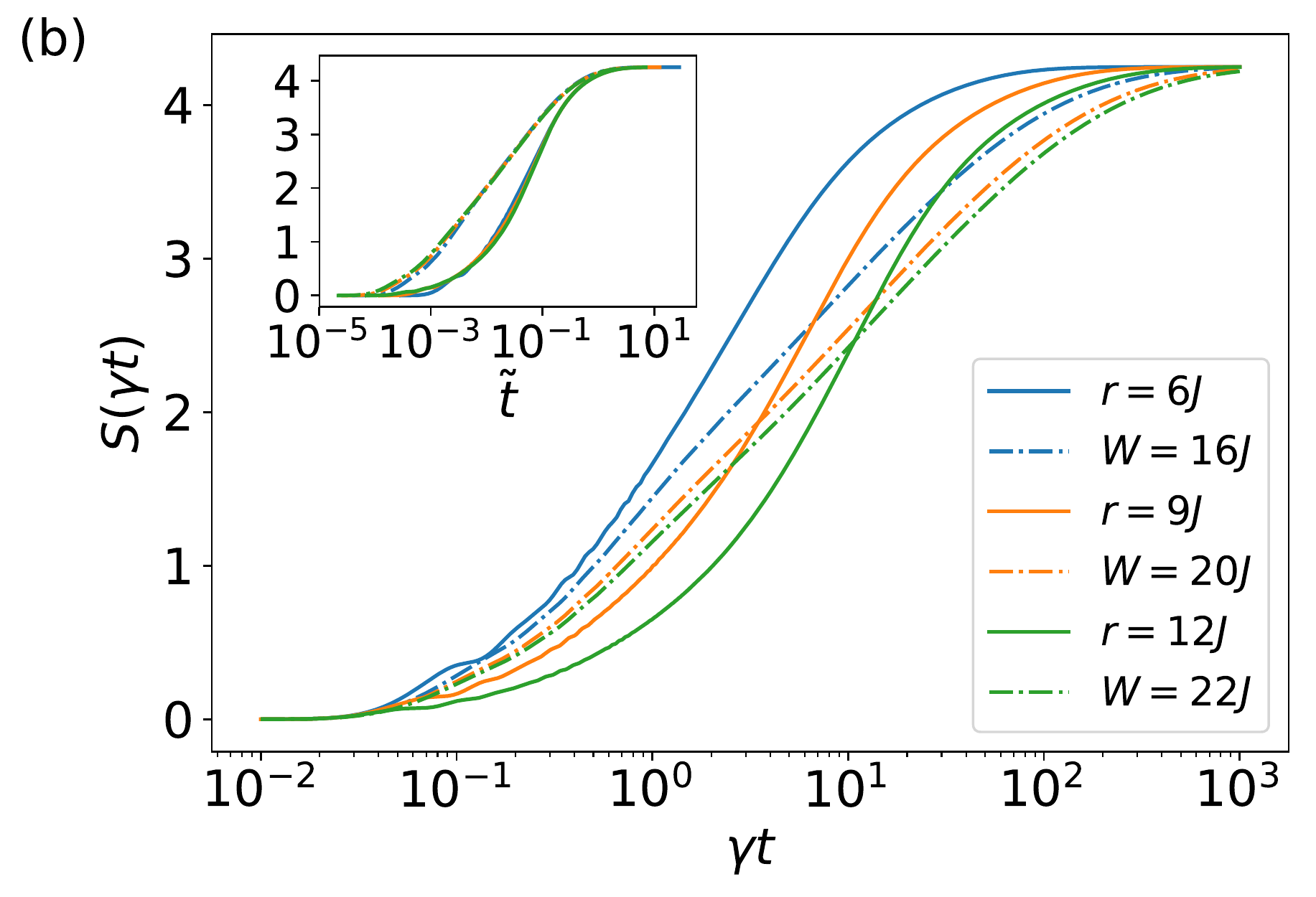}}
     {\includegraphics[width=0.67\columnwidth]{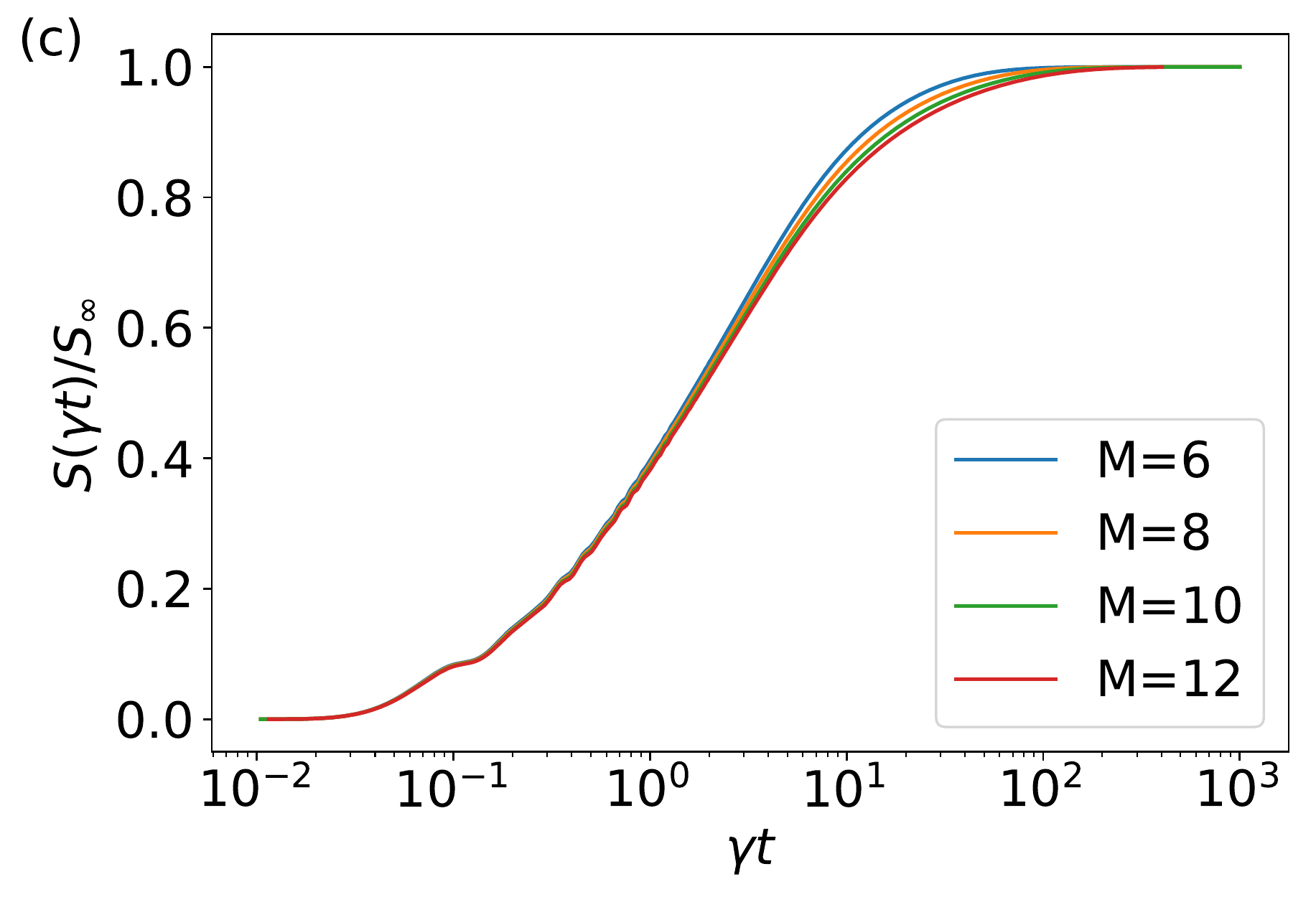}}
    \caption{Growth of the von Neumann entropy $S(t)$ in time for a chain at half filling from the initial
charge-density-wave state.
    In (a), the dotted lines are the results for $V=0$, and the solid lines are the results for $V=J$. The inset figure shows the entropy as a function of the scaled time $\tilde{t}$. In (b), the entropies for the Stark-localized system and the disorder-localized system are compared. 
 For the disorder model, the on-site potential is $W_i = w_i$, with $w_i$ being a random number uniformly distributed in the interval $[-W, W]$. The results  are averaged over $100$ disorder realizations. \added{The inset shows the entropy as a function of the scaled time $\tilde{t}$. For the disorder system, the scaled time is defined as $\tilde{t} = \gamma t (J/W)^2$~\cite{Fischer2016PhysRevLett.116.160401}.}
  The system size for (a) and (b) are $M=8$. (c) shows the normalized entropy for different system sizes at $r=6J$. The interaction strength for (b) and (c) is $V=J$.
    The dissipation rate is $\gamma=0.1J$. 
    }\label{St}
\end{figure*}


\added{Let us now investigate the role of interactions. }
For a large potential gradient $r$ with $VJ/r^2 \ll 1$, based on the perturbation theory to the leading order in $V$~\citep{Fischer2016PhysRevLett.116.160401},
the dynamics of the mean occupations in the eigenbasis is approximately governed by
\begin{eqnarray}\label{nkV}
\langle \dot{{\tilde n}}_k \rangle &=& \gamma \sum_q R_{kq}( \langle \tilde{n}_q \rangle- \langle\tilde{n}_k\rangle) \notag\\
&&+ \gamma \left(\epsilon \frac{VJ}{r^2}\right)^2  \left({\langle \tilde{n}_{k+1}\rangle+\langle \tilde{n}_{k-1}\rangle-2\langle \tilde{n}_{k}\rangle}\right)^2.
\end{eqnarray}
Here $\epsilon$ is a numerical constant
of order $1$, \added{which is adjusted to optimize the matching of the approximated results with the exact ones~\footnote{{For the edge sites, we take $\langle \dot{{\tilde n}}_1 \rangle = \gamma R_1 \left(\langle \tilde{n}_{2} \rangle - \langle \tilde{n}_{1} \rangle\right) - \gamma \left(\epsilon \frac{VJ}{r^2}\right)^2 \left(\langle \tilde{n}_{2} \rangle - \langle \tilde{n}_{1} \rangle\right)^2 $, and $\langle \dot{{\tilde n}}_M \rangle = \gamma R_1 \left(\langle \tilde{n}_{M-1} \rangle - \langle \tilde{n}_{M} \rangle\right) - \gamma \left(\epsilon \frac{VJ}{r^2}\right)^2 \left(\langle \tilde{n}_{M-1} \rangle - \langle \tilde{n}_{M} \rangle\right)^2$}}.}
As shown in Fig.~\ref{IVt}, the long-time behavior of the imbalance (orange solid line) is well captured by
the simple equations in \eqref{nkV}, whose corresponding result is shown in black dot-dashed line.
Note that the first term in Eq.~\eqref{nkV} is identical to Eq.~\eqref{nk} for the non-interacting
case. The second term describes the contribution from interactions, which leads to an interaction-assisted-hopping on the order of $\sim (VJ/r^2)^2$.  It implies that interactions will enhance the decay of the imbalance. This is confirmed in Fig.~\ref{IVt}, where we see that
a strong interaction enhances the decay of the imbalance at short times and then leads to a stretched exponential decay before approaching the steady-state value.
\added{The inset in Fig.~\ref{IVt}(b) shows the stretching exponent $\beta$ as a function of the interaction strength $V$ (see Fig. S3 in Supplementary Material for more details of the curve fitting).}

%

As a second quantity, we now study the von Neumann entropy of the \added{whole} system,
\begin{equation}
S(t) = - {\rm Tr}\left\{ \rho(t) \log[\rho(t)]\right\},
\end{equation}
which quantifies the heating induced by the bath.
Figure~\ref{St} shows the time evolution of the entropy for a chain at half filling from the initial charge-density-wave state. From Fig.~\ref{St}(a) we can see that
the rate of entropy growth is set by the field strength $r$. As shown in the inset of  Fig.~\ref{St}(a), the entropies for different $r$ collapse onto each other as a function of the scaled time $\tilde{t}$. By comparing the results for the non-interacting case (dotted lines) with those for the interacting case with $V=J$ (solid lines), we find that the effect of  interactions is weak and tends to enhance the growth of entropy.

\begin{figure}[!htbp]
    \centering
{\includegraphics[width=0.50\columnwidth]{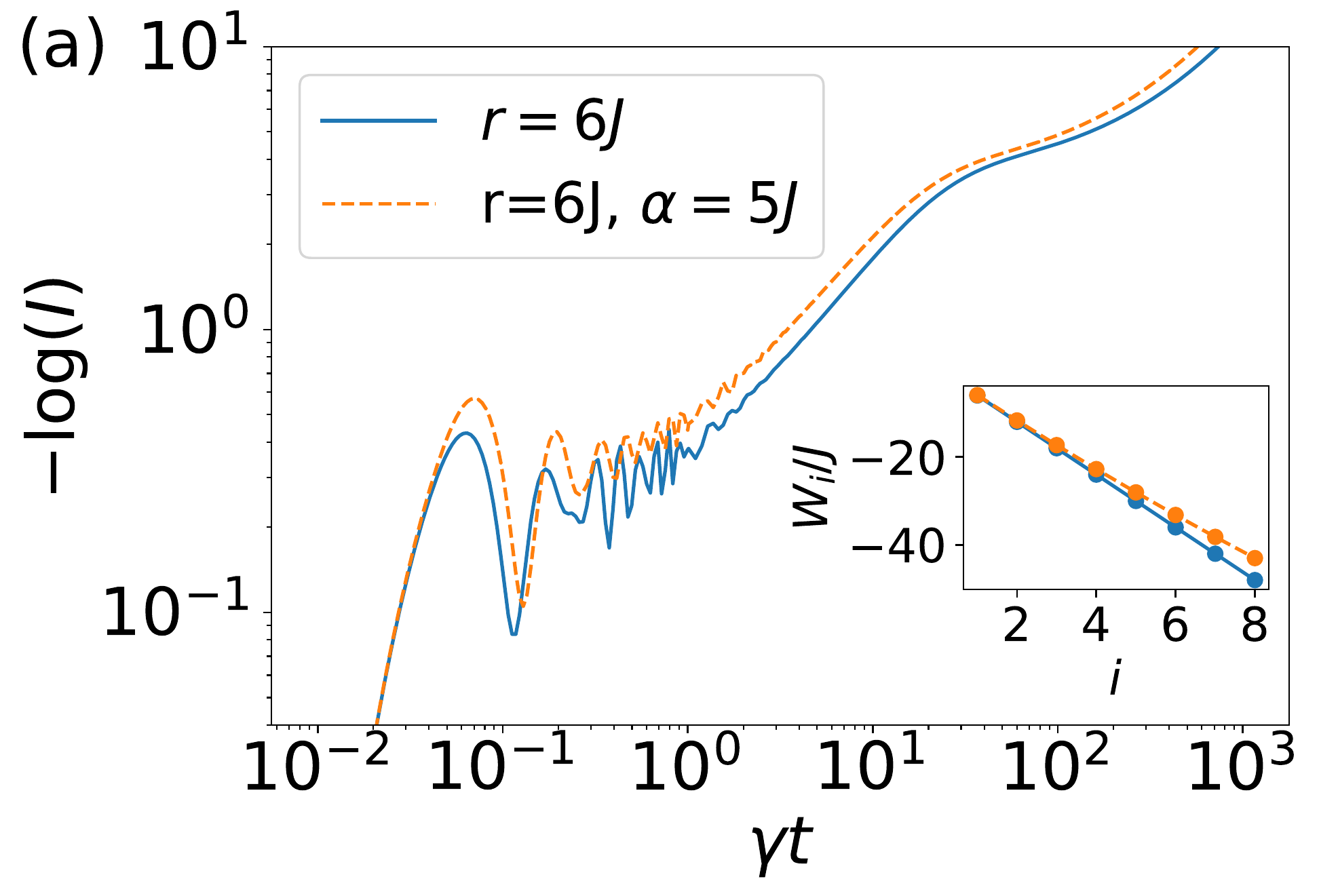}}
{\includegraphics[width=0.48\columnwidth]{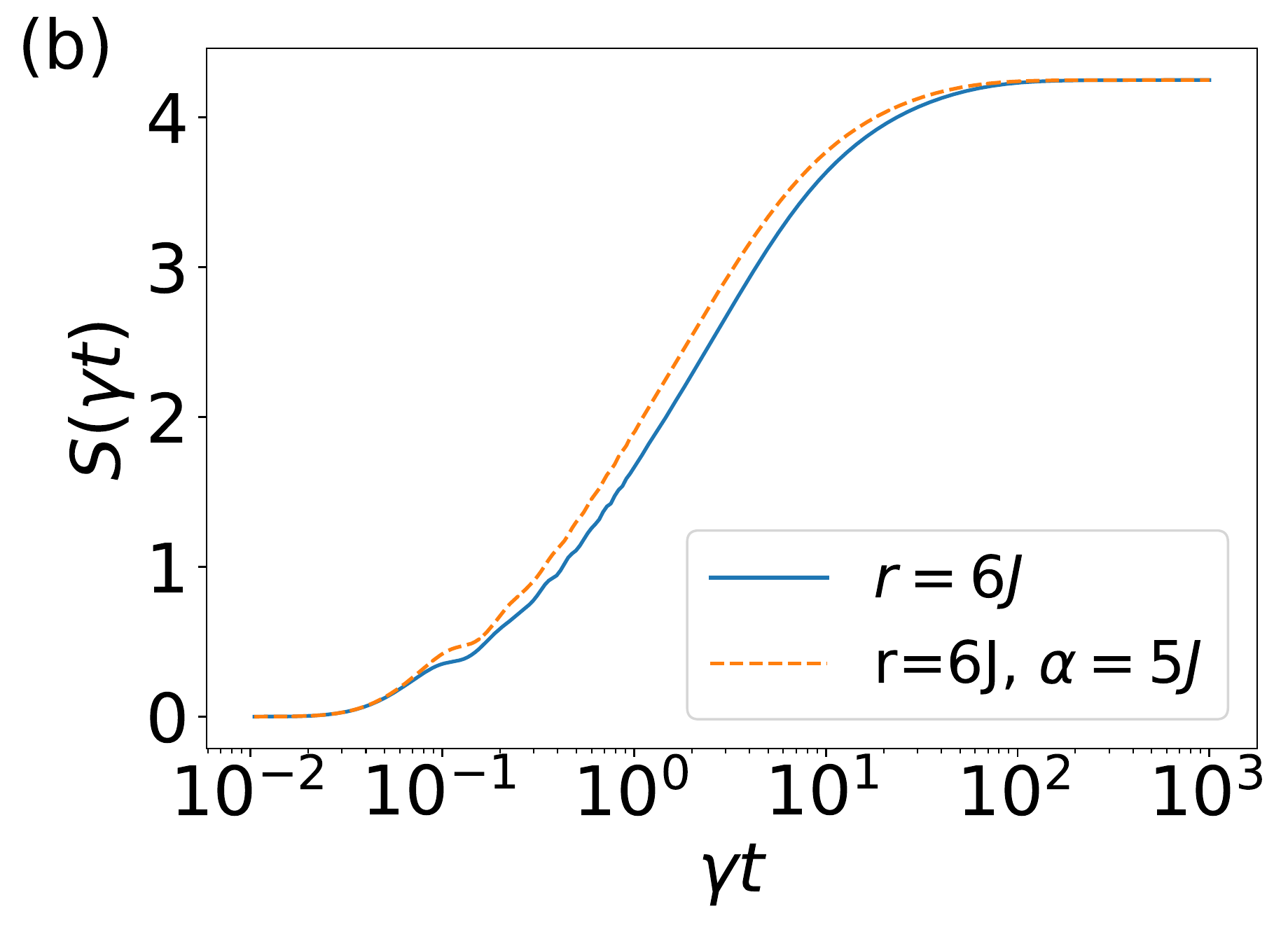}}
    \caption{\added{Comparison of dynamics of the imbalance (a) and entropy (b) from the initial charge-density-wave state for different on-site potentials $W_i$ (inset). The blue solid line is the result for a linear potential with $W_i = -ri$. The orange dashed line depicts the result for a quadratic potential with $W_i = -ri+\alpha(i/M)^2$. The parameters are $M=8$, $\gamma=0.1J$, $V=J$.}
    }\label{alpha}
\end{figure}

One of our main findings is shown in Fig.~\ref{St}(b), where we compare the entropy growth of the Stark-localized system (solid lines) to that of a disorder-localized system (dot-dashed lines) in a semilog plot. 
\added{For the latter, the entropy exhibits logarithmically slow growth, with a linear slope in the semilog plot found in a wide time window covering about two decades~\cite{Levi2016PhysRevLett.116.237203}.
In contrast, we do not find such an extended region with a linear slope for the Stark localized system: a linear slope is found only at the inflection point associated with the on-set of saturation.
This observation is robust to field strength, as is supported by the inset, which shows the collapse of entropies for various field strengths at large scaled times $\tilde t$.
It is also not related to the particular parameters selected in the plot, such as coupling rate $\gamma$ and interaction strength $V$, whose impacts on the dynamics are found to be weak (see Figs. S4 and S5 of the Supplementary Material for more details).} In Fig.~\ref{St}(c), we plot the time evolution of the entropy (normalized by its maximum, i.e. infinite-temperature value $S_{\infty} = \log\{M!/[(M/2)!]^2\}$) for different system sizes $M$. We find a very weak dependence on the system size only \added{and no indication that the linear slope near the inflection point starts to extend over a larger time interval with increasing $M$.

Note that the observed non-logarithmic behavior is also not associated with the exceptional behavior found for a purely linear potential gradient in the closed system~\cite{schulz2018stark}. Namely it was shown that a purely linear potential is not enough for the Stark system to exhibit generic MBL behavior, which only occurs when, e.g., a small quadratic potential is added. Such a big difference brought by the additional weak field is absent in the open system. As shown in Fig.~\ref{alpha},
adding such a potential to our open system does not lead to a qualitative change in the dynamics of both population imbalance (a) and entropy (b).}

In conclusion, we investigate the relaxation dynamics of an open chain of interacting spinless
fermions in the presence of a strong electric field coupled to a dephasing bath.
The closed (isolated) system is shown by previous studies~\cite{van2018bloch,schulz2018stark} to
exhibit non-ergodic behavior analogous to conventional disorder-induced MBL system. However, when coupled to a dephasing bath, the Stark system shows qualitatively different relaxation dynamics towards steady state.
\added{We show that in contrast to a disordered system~\cite{Fischer2016PhysRevLett.116.160401}, the decay of the population imbalance is described by a stretched exponential only in the presence of interactions.}
Another stark difference is the fact that the growth of the von Neumann entropy is not logarithmically slow, as it was found for the disordered system~\cite{Levi2016PhysRevLett.116.237203}.
Our findings can immediately be tested experimentally with ultracold atoms in optical lattices by employing the techniques of Ref.~\cite{Bloch2017PhysRevX.7.011034}, where the impact of a dephasing bath on the decay of quasi-disorder-induced MBL was investigated.
\added{In such an experiment, one would rather consider spinful fermions with on-site interactions (the numerical treatment of which is more difficult as a result of the enlarged state space and beyond the scope of this paper). However, also in this case qualitative differences in the relaxation dynamics of the disorder localized and the Stark localized open system can be expected, since we observed profound differences already in the limit of vanishing interactions.
}

\begin{acknowledgments}
We acknowledge discussions with Markus Heyl.
This research was funded by the Deutsche
Forschungsgemeinschaft (DFG) via the Research Unit
FOR 2414 under Project No. 277974659.
\end{acknowledgments}

\bibliography{ref}

\pagebreak

\clearpage

\onecolumngrid
\begin{center}
  \textbf{\large Supplementary Material for\\``Bath-induced decay of Stark many-body localization"}\\[.2cm]
  Ling-Na Wu and Andr{\'e} Eckardt\\[.1cm]
  {\itshape Max Planck Institute for the Physics of Complex Systems, D-01187, Dresden}
\end{center}

\setcounter{equation}{0}
\setcounter{figure}{0}
\setcounter{table}{0}
\setcounter{page}{1}
\renewcommand{\theequation}{S\arabic{equation}}
\renewcommand{\thefigure}{S\arabic{figure}}
\renewcommand{\bibnumfmt}[1]{[S#1]}

\section{Solution to Eq.~(6) in the main text}\label{mean-occupation}
To solve Eq.~(6) in the main text, we rewrite it in matrix form as
\begin{equation}\label{AB}
{\dot {\bf n}} = \gamma R_1A {\bf n},
\end{equation}
where ${\bf n} = (\langle \tilde{n}_1 \rangle, \ldots, \langle \tilde{n}_M \rangle)^T$, and $A$ is a Tridiagonal quasi-Toeplitz  matrix
\begin{equation}
A = \left(
\begin{array}{ccccc}
-1 & 1 & 0&\cdots& 0\\
1 & -2 & 1& &\vdots \\
0& \ddots & \ddots & \ddots &0 \\
\vdots& & 1 &  -2&  1 \\
0 &  \cdots& 0 & 1&-1
\end{array}
\right),
\end{equation}
with eigenvalues $a_k = -2+2\cos(\frac{(k-1)\pi}{M})$ ($k=1,\ldots,M$) and eigenvectors $(u^{(k)}_1,u^{(k)}_2, \ldots ,u^{(k)}_M)^T$ with
$u^{(k)}_q = \sqrt{\frac{2-\delta_{k,1}}{M}} \cos\left(\frac{(k-1)(2q-1)\pi}{2M}\right)$~\cite{yueh2005eigenvalues}.
Its solution is given by
\begin{eqnarray}\label{na}
{\langle \tilde{n}_k \rangle} &=&\sum\limits_{q,p}{ f_k(q,p) e^{\gamma R_1 t [-2 + 2\cos(\frac{(q-1)\pi}{M})]}\langle \tilde{n}_p(0)\rangle},
\end{eqnarray}
with $f_k(q,p) = \frac{2-\delta_{q,1}}{M}  \cos\left(\frac{(q-1)(2k-1) \pi}{2M}\right) \cos\left(\frac{(q-1)(2p-1)\pi}{2M}\right)$.

\begin{figure}[!htbp]
    \centering
{\includegraphics[width=0.5\columnwidth]{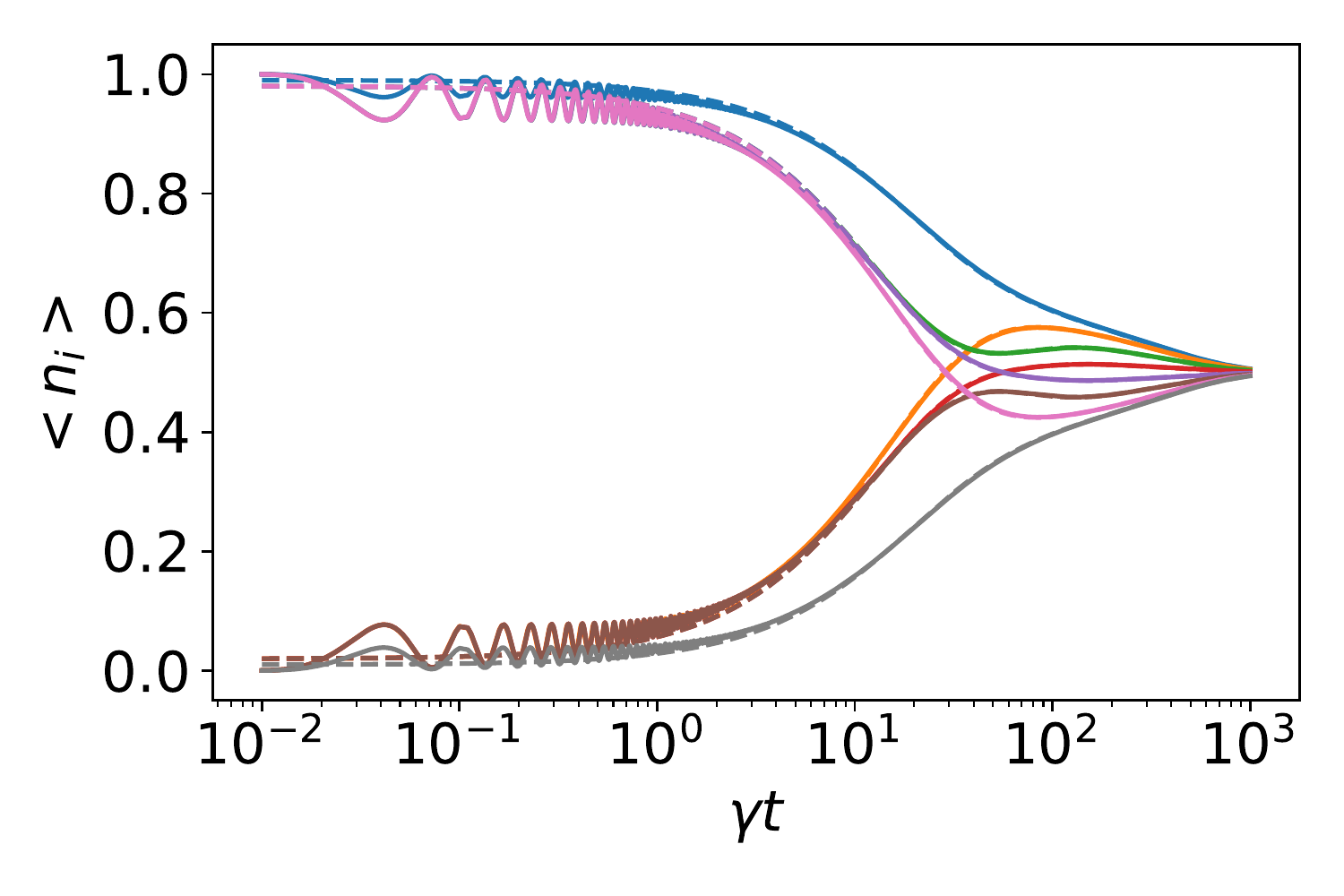}}
 \caption{Time evolution of the mean occupation $\langle n_i \rangle$ on site $i$. The solid lines are the exact results from numerical integration of the master equation (2) in the main text, and the dashed lines denote the approximated results from Eq.~(\ref{na}) by using $\langle n_i\rangle = \sum_k |\psi_{ik}|^2 \langle \tilde{n}_k \rangle$.
    The parameters are, number of sites $M=8$, coupling rate $\gamma=0.1J$, field gradient $r=10J$, interaction strength $V=0$.
    }\label{Sapp}
\end{figure}

In Fig.~\ref{Sapp}, we compare the mean occupation from numerical integration of the master equation (2) in the main text (solid lines) and that by using the approximated result of Eq. (\ref{na}) (dashed lines). Except for the initial coherent oscillations, which are absent in the latter due to the neglect of the off-diagonal terms in the density matrix, Eq.~(\ref{na}) well describes the time evolution of the mean occupation.

\section{Stretched exponential fitting}
\begin{figure}[!htbp]
    \centering
{\includegraphics[width=0.9\columnwidth]{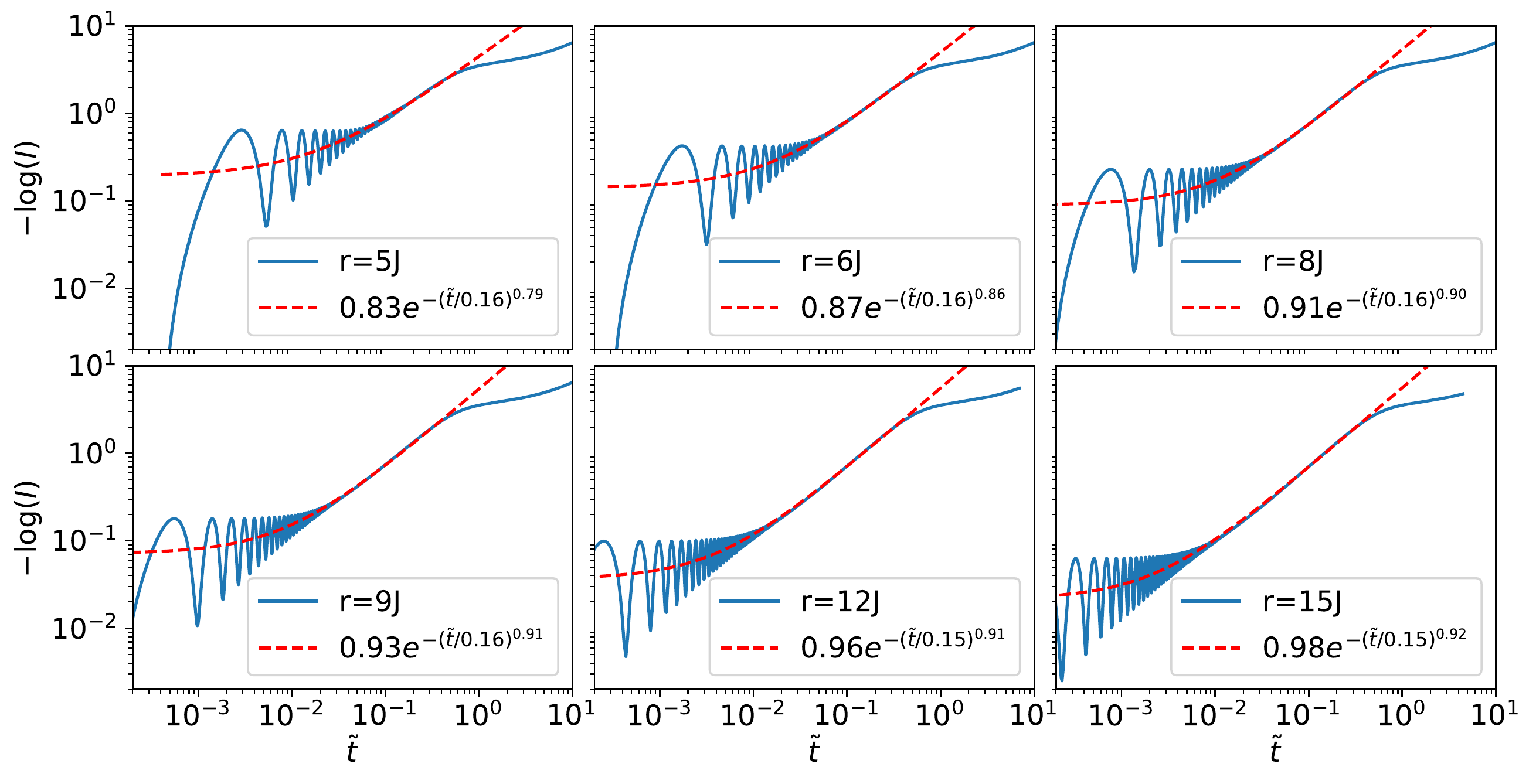}}
 \caption{Dynamics of the imbalance for noninteracting systems with different field strengths $r$ from the initial charge-density-wave state. The red dashed lines are fitting curves based on stretched exponential function $I=I_0 e^{-(\tilde{t}/t_0)^\beta}$.
 The scaled time $\tilde{t} = \gamma t (J/r)^2$.
    The parameters are $M=8$, $\gamma=0.1J$, $V=0$.
    }\label{IV0}
\end{figure}

\begin{figure}[!htbp]
    \centering
{\includegraphics[width=0.9\columnwidth]{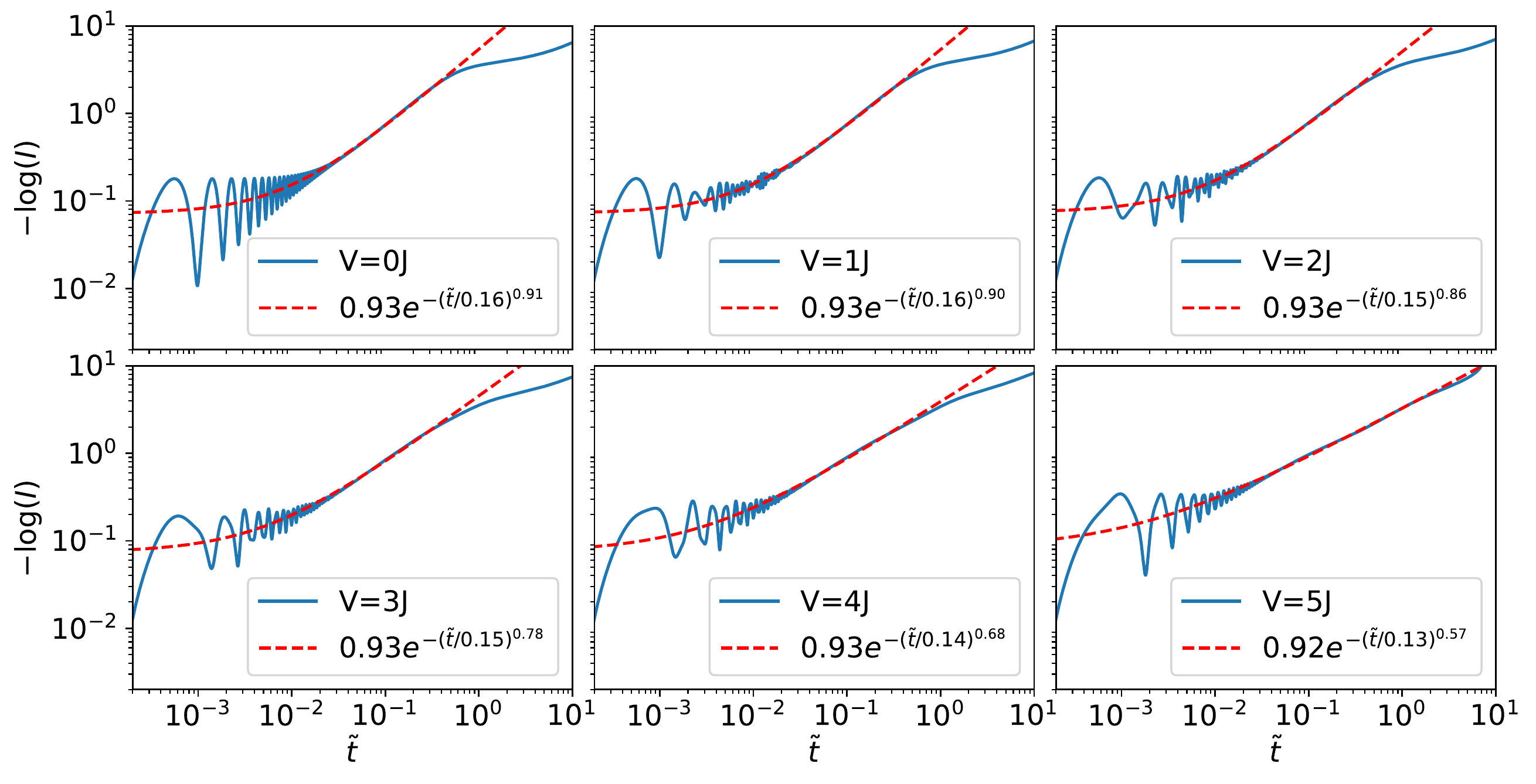}}
 \caption{Dynamics of the imbalance for interacting systems with different interaction strengths $V$ from the initial charge-density-wave state. The red dashed lines are fitting curves.
    The parameters are $M=8$, $\gamma=0.1J$, $r=9J$.
    }\label{fit_ISV}
\end{figure}

Fig.~\ref{IV0} shows the dynamics of the imbalance for noninteracting systems with different field strengths $r$ from the initial charge-density-wave state.
The red dashed lines are fitting curves based on stretched exponential function $I=I_0 e^{-(\tilde{t}/t_0)^\beta}$. The data with $\tilde t<0.5$ are used for the fitting.
Fig.~\ref{fit_ISV} shows similar results for different interaction strengths $V$.

\section{Dependence of relaxation dynamics on system parameters}
In this section, we study the dependence of relaxation dynamics on various system parameters for a large field gradient ($r=15J$).
\subsection{Dependence on coupling rate $\gamma$}
\begin{figure}[!htbp]
    \centering
{\includegraphics[width=0.45\columnwidth]{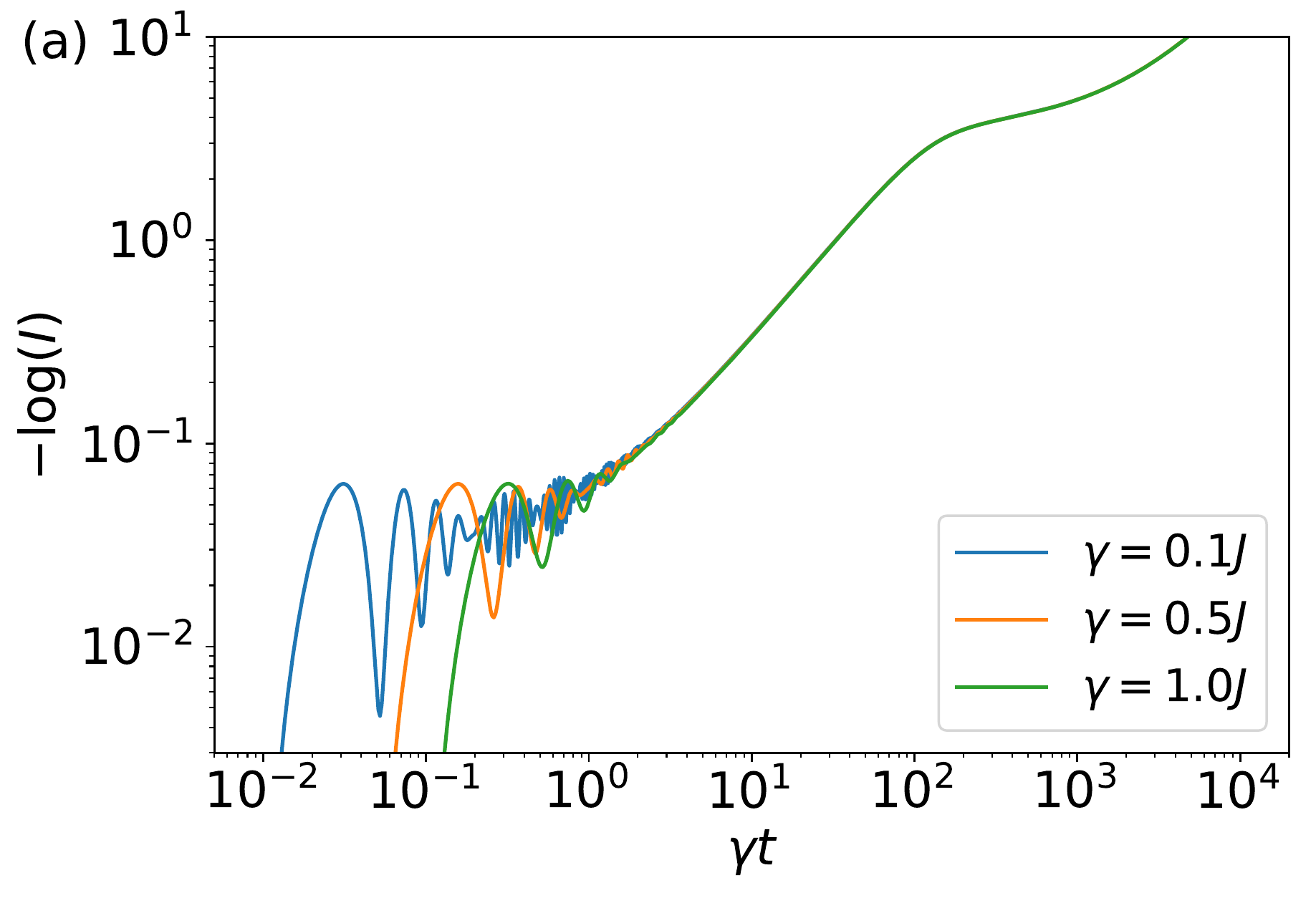}}
{\includegraphics[width=0.45\columnwidth]{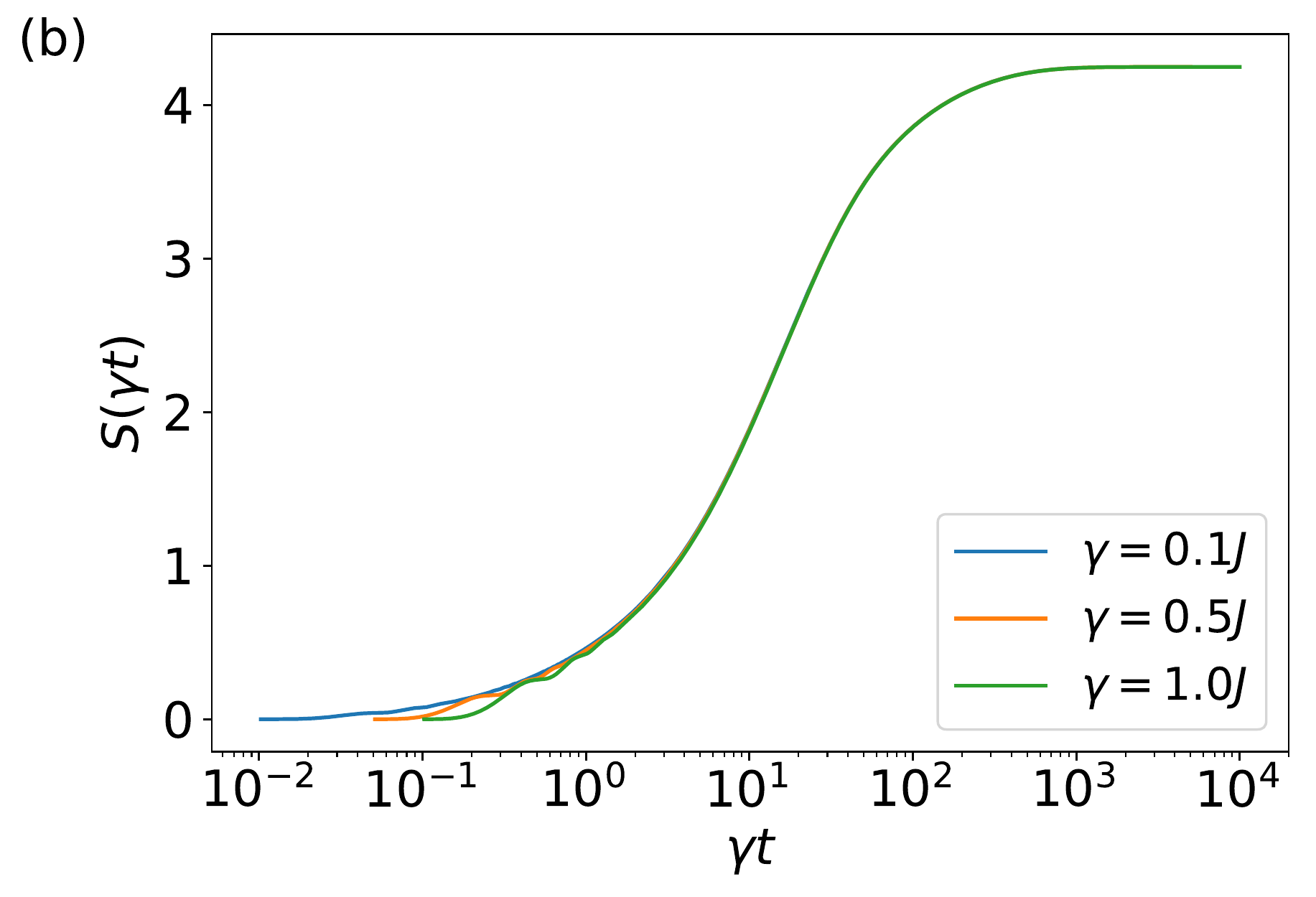}}
 \caption{Dynamics of the imbalance (a) and entropy (b) for different system-bath coupling rates $\gamma$ from the initial charge-density-wave state.
    The parameters are $M=8$, $V=J$, $r=15J$.
    }\label{ISg}
\end{figure}

Fig.~\ref{ISg} shows the dynamics of the imbalance (a) and entropy (b) for different weak system-bath coupling rates $\gamma$ with $\gamma/J  \lesssim  1$.
At short times $\gamma t \lesssim 1$, the coupling rate $\gamma$ sets the oscillation rate of the dynamics. While for long time evolution with $\gamma t \gg 1$, the dynamics exhibits a collapse when rescaling the time axis by $\gamma$.

\subsection{Dependence on interaction strength $V$}
\begin{figure}[!htbp]
    \centering
{\includegraphics[width=0.45\columnwidth]{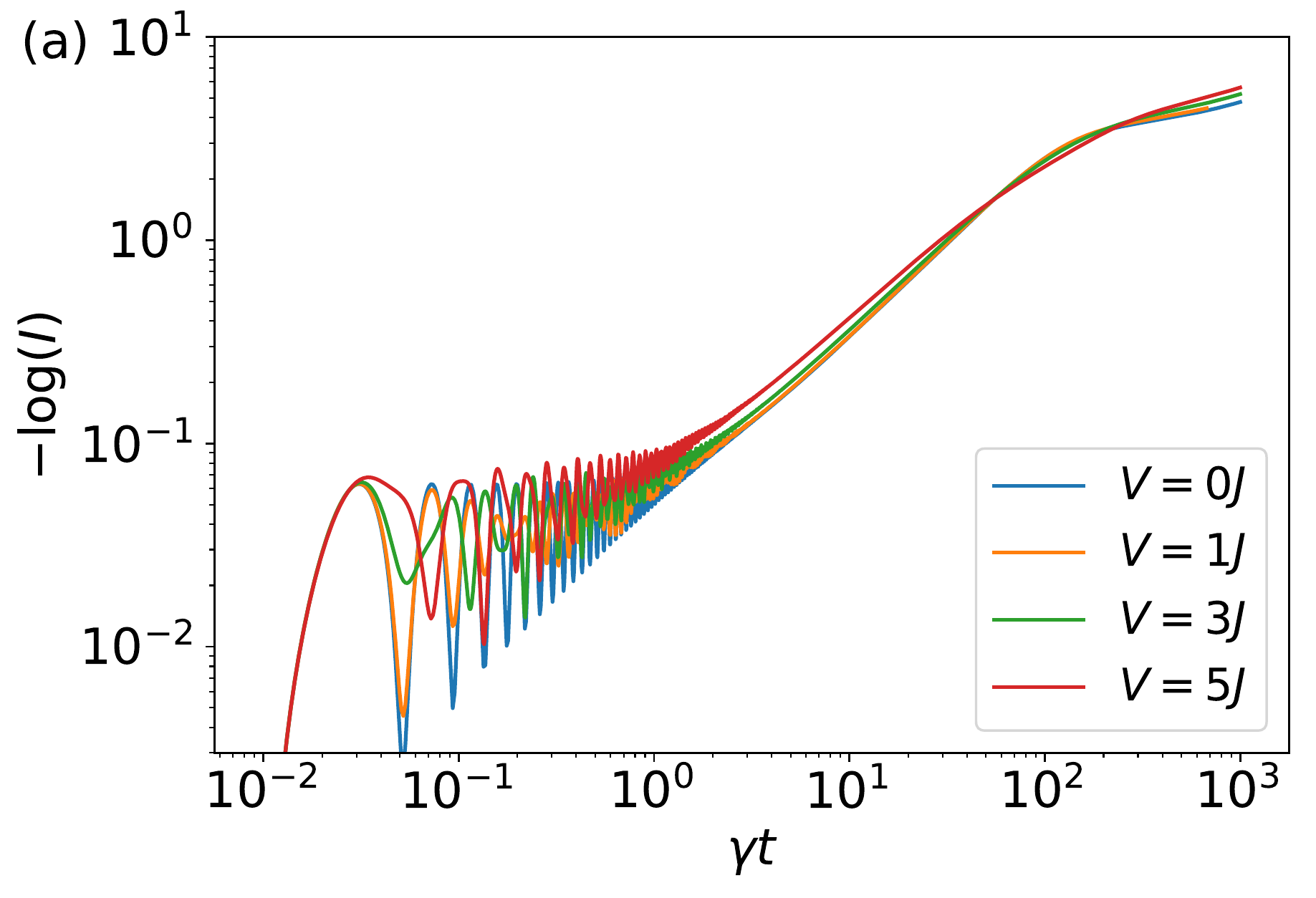}}
{\includegraphics[width=0.45\columnwidth]{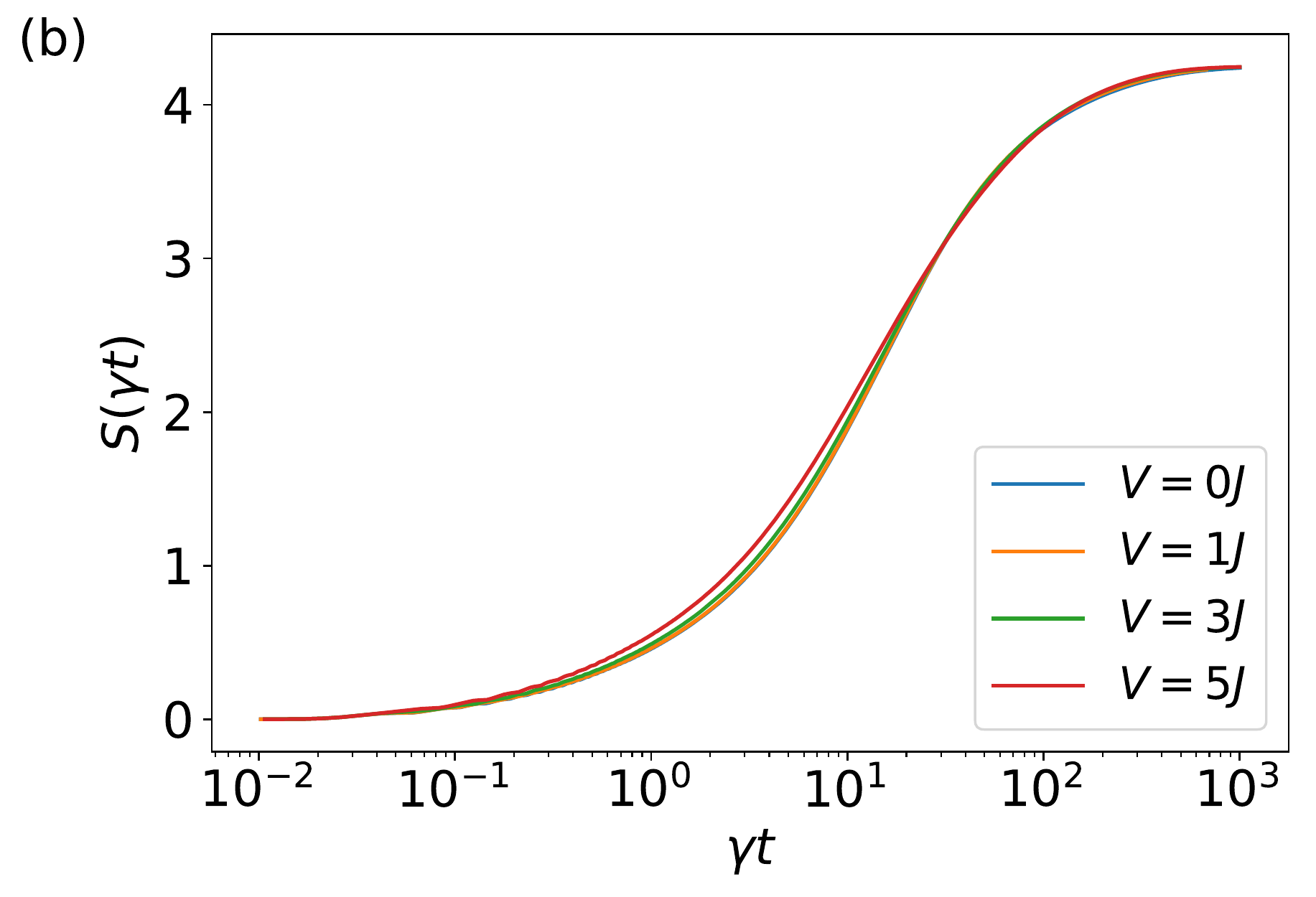}}
 \caption{Dynamics of the imbalance (a) and entropy (b) for different interaction strengths $V$ from the initial charge-density-wave state.
    The parameters are $M=8$, $\gamma=0.1J$, $r=15J$.
    }\label{ISV}
\end{figure}

Fig.~\ref{ISV} shows the dynamics of the imbalance (a) and entropy (b) for different interaction strengths $V$.
The effect of interaction is found to be subleading compared to dephasing noise.

\end{document}